\documentclass[aps,pra,twocolumn,groupedaddress,showpacs,floatfix,superscriptaddress]
{revtex4}
\usepackage{graphicx}
\usepackage{bm}
\usepackage{amsmath}
\usepackage{amssymb}
\usepackage{latexsym}
\usepackage{epsfig}
\usepackage{amsbsy}
\usepackage{array}
\usepackage{amssymb}
\usepackage{setspace}
\usepackage{subfigure}
\usepackage{epstopdf}
\usepackage[colorlinks,urlcolor=blue,linkcolor=blue,citecolor=blue]{hyperref}
\usepackage[english]{babel}
\usepackage{float}
\usepackage{calc}
\usepackage{color}
\usepackage{braket}
\usepackage{dcolumn}

\begin{document}

\preprint{Physical Review Letter}

\title{Suppressing laser phase noise in an optomechanical system}

\author{Yexiong Zeng}
\affiliation{School of physics, Dalian university of technology, Dalian  116024, China}

\author{Biao Xiong}
\affiliation{College of Physics and Electronic Science, Hubei Normal University, Huangshi 435002, People’s
Republic of China}
\author{Chong Li\footnote{Corresponding author  lichong@dlut.edu.cn}}
\affiliation{School of physics, Dalian university of technology, Dalian  116024, China}


\begin{abstract} 
We propose a scheme to suppress the laser phase noise without increasing the optomechanical single-photon coupling strength. In the scheme, the parametric amplification terms, created by Kerr and Duffing nonlinearities, can restrain laser phase noise and strengthen the effective optomechanical coupling, respectively. Interestingly, decreasing laser phase noise leads to increasing thermal noise, which is inhibited by bringing in a broadband-squeezed vacuum environment.  To reflect the superiority of the scheme, we simulate quantum memory and stationary optomechanical entanglement as examples, and the corresponding numerical results demonstrate that the laser phase noise is extremely suppressed.  Our method can pave the way for studying other quantum phenomena.
 \end{abstract}


\maketitle

\section{ Introduction}
Cavity optomechanics has sparked extensive theoretical and experimental research interest in the last decade due to its various applications~\cite{RevModPhys.86.1391,Kippenberg1172,Naik2006,Sankey2010,Park2009,Rodgers2010,PhysRevX.1.021011,PhysRevX.8.011031,PhysRevX.5.041024,PhysRevLett.119.053601,Liu2018,Zhong2018}: detecting weak-force, mass, displacements, and orbital angular momentum~\cite{doi:10.1063/1.2012461,PhysRevX.6.021001,PhysRevLett.108.120801,Zhang2020}; creating macroscopic nonclassical states~\cite{PhysRevLett.116.163602,PhysRevA.89.014302}; achieving mechanical squeezing~\cite{Wollman952,Xiong2020151}; obtaining optical nonreciprocity \cite{Yan2019}.
These advancements exhibit potential advantages of cavity optomechanics in quantum metrology~\cite{PhysRevLett.104.213603}, quantum information processing~\cite{Rabl2010,PhysRevA.91.053854,PhysRevA.100.043835,PhysRevA.101.013802,Jing2015,Jing2017,Zeng:20,doi:10.1063/5.0035498,L2013,PhysRevA.90.053833,Zhao2019,PhysRevA.89.023849,PhysRevA.91.033835}, and fundamental physics questions~\cite{PhysRevA.99.063811,doi:10.1002/andp.201900596,Xiong:18,PhysRevA.102.023707,PhysRevA.92.033806}.

Though optomechanical systems have brought wide attention, the single-photon optomechanical coupling $g_0$ is small in experiment~\cite{Liao_2014,PhysRevA.101.063802}. People usually bring in classical driving lasers to improve the effective optomechanical coupling strength~\cite{PhysRevLett.110.153606,PhysRevA.91.033818,PhysRevA.94.053807,PhysRevA.95.063825,PhysRevA.98.053802}.  However, the phase and amplitude of the driving laser have small fluctuation — the so-called laser phase and amplitude noise. Schliesser et al. firstly observed the laser phase noise in experiment~\cite{Schliesser2008}. Generally, laser amplitude noise can be approximately neglected by stabilizing the laser intensity~\cite{PhysRevA.83.063838,PhysRevA.94.063636}. Recently, many efforts have been devoted to study the influence of laser phase noise on quantum physics: ground-state cooling~\cite{PhysRevA.78.021801,PhysRevA.80.033821,Farman:13,PhysRevA.80.063819,PhysRevLett.123.153601,PhysRevLett.118.233604}; quantum state estimation~\cite{PhysRevLett.114.223601}; weak-force sensing~\cite{PhysRevA.98.053841,Mehmood_2019}; squeezing mechanical oscillators~\cite{Gu:20} and output light of an optical cavity~\cite{PhysRevA.89.033810}; quantum memory~\cite{PhysRevA.91.033828}; optomechanical entanglement~\cite{PhysRevA.84.032325,PhysRevA.84.063827,Ahmed_2019}. Many of these studies demonstrate that phase noise has destructive effects on quantum phenomena.
Optomechanical entanglement is a macroscopic quantum phenomenon and is significant for quantum information processing\cite{PhysRevA.96.053831,Yan:19}. Some works have discussed the influence of the Kerr nonlinear medium on the stationary optomechanical entanglement in the presence of laser phase noise~\cite{Zhang_2013,Zhang_20131,Ahmed_2019}. Although they claim that Kerr nonlinearity can promote entanglement to some extent, they leave an unresolved contradiction: strong optical force coupling accompanied by huge laser phase noise. Specifically, the effect of laser phase noise is described by $\sqrt{2}\alpha\dot{\varphi}(t)$ where $\alpha$ and $\dot{\varphi}(t)$ are the mean value of intracavity field and time derivatives of phase noise, respectively. The effective optomechanical coupling is $\sqrt{2}g_0\alpha$ where the single-photon coupling $g_0$ is very weak in experiment.  Therefore, people usually improve the effective optomechanical coupling $\sqrt{2}g_0\alpha$  by raising $\alpha$. However, the increase of $\alpha$ is unavoidable to enlarge the effect of phase noise. Naturally, we  raise  a  novel and interesting idea:  Can we simultaneously strengthen the effective optomechanical coupling and inhibit the effect of laser phase noise? 

In this paper, we present a scheme to improve the effective optomechanical coupling and inhibit laser phase noise at the same time. 
The system consists of an optical cavity, a mechanical oscillator, and a Kerr nonlinear medium. We obtain the effective Hamiltonian with optical and mechanical parametric amplification terms after linearizing the systemic Hamiltonian and then transform the system into the squeezing frame. According to the analytical results, we found that the effective optomechanical coupling is enhanced, and the laser phase noise decreases exponentially by adjusting the squeezing parameters. Generally, we can introduce a broadband-squeezed vacuum environment to suppress the increased thermal noise around the squeezed cavity field and mechanical oscillator. However, the enhanced thermal noise around the mechanical oscillator has little influence on the system due to its tiny decay, which means we only need to suppress the enlarged thermal noise around the squeezed cavity field by exploiting the vacuum environment. Our calculation shows that our proposal effectively solves the contradiction:  improving effective optomechanical coupling leads to enlarging laser phase noise. To display our scheme, we exploit it to simulate quantum memory and stationary optomechanical entanglement as examples. Our numerical results show our strategy can significantly protect both the quantum memory and the stationary optomechanical entanglement.  

This paper is organized as follows: in Sec.~\ref{Physical Model},  We explain the theoretical model, derive the effective Hamiltonian, and obtain the dynamic equation of the additional mode describing the effect of phase noise. We demonstrate some actual physical phenomena (quantum memory and stationary optomechanical entanglement)  in Sec.~\ref{Demonstrating some actual quantum phenomenon}.  Finally, we give a conclusion in Sec.~\ref{Conclusion}.

\section{Physical Model}
\label{Physical Model}
\begin{figure}[tbp]
\centering
{\includegraphics[width=0.9\columnwidth]{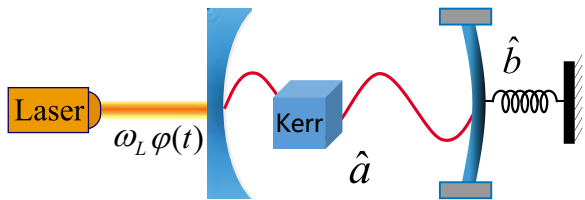}}
\caption{Schematic illustration of an optomechanical system. The system includes cavity mode $\hat{a}$ and mechanical mode $\hat{b}$. A Kerr nonlinearity medium is located in the cavity that is driven by a classical laser. The system includes a Duffing nonlinearity related to the mechanical mode. $\omega_L$ and $\varphi(t)$ denote the frequency and the phase fluctuation of the optical driving laser, respectively.}
\label{figure1}
\end{figure}

We consider an optomechanical system and its schematic diagram is illustrated in Fig.~\ref{figure1}. In this setup, a classical laser drives the optical cavity, and  a Kerr medium is located in the cavity.  Kumar et al. have shown that Kerr medium inside an optomechanical system can effectively inhibit the normal mode splitting~\cite{PhysRevA.81.013835}.  In the rotating frame with frequency $\omega_L$, the Hamiltonian of the system can be written as ($\hbar=1$)~\cite{PhysRevA.101.023841,PhysRevA.99.013843}
\begin{equation}
\begin{split}
\hat{H}_{\text{rot}}=&\Delta_0\hat{a}^{\dagger} \hat{a}+\omega_{m}\hat{b}^{\dagger}\hat{b}-g_{0} \hat{a}^{\dagger} \hat{a} (\hat{b}^{\dagger}+\hat{b})+u\hat{a}^{\dagger 2}\hat{a}^2
\\
&-\frac{\eta}{2}(\hat{b}+\hat{b}^{\dagger})^4+iE_{L}\left(e^{i \varphi(t)}\hat{a}^{\dagger}-e^{-i \varphi(t)}\hat{a}\right),
\end{split}
\label{eq1}
\end{equation}
where $\hat{a}$ ($\hat{a}^{\dagger}$) and $\hat{b}$ ($\hat{b}^{\dagger}$) are the annihilation (creation) operators of the cavity field and the mechanical oscillator, respectively. The optical cavity mode with resonance frequency $\omega_c$ contains a  Kerr nonlinear medium with Kerr coefficient $u$. The mechanical oscillator with resonance frequency $\omega_m$ is accompanied by a Duffing  nonlinear  term with amplitude $\eta$ and coupled to  the cavity field with coupling strength  $g_0$. By coupling the mechanical mode to a qubit, a strong Duffing nonlinearity can be obtained, in which the nonlinear amplitude $\eta$ can reach $10^{-4}\omega_m$\cite{PhysRevA.91.013834}.  $\omega_L$ is the frequency of the classical driving laser, $\varphi(t)$ describes the laser phase noise—a zero-mean stationary Gaussian stochastic process, and $E_L$ is the strength of the driving laser. Generally, one can adjust the driving strength  $E_L=\sqrt{2\kappa P_s/\hbar \omega_L}$ by controlling the  input power $P_s$ where $\kappa$ is the decay of the cavity through its input port. The amplitude noise of the driving laser is negligible compared to the phase noise via stabilizing laser source, so we can neglect the amplitude noise of the driving laser in this paper~\cite{PhysRevA.84.032325,PhysRevA.91.033828}. 

The optical Kerr effect has been studied theoretically and experimentally, and  the corresponding nonlinear Kerr coefficient has the following form~\cite{Brasch357,PhysRevA.80.065801} 
\begin{equation}
u=\frac{\hbar \omega_{c}^{2} c n_{2}}{n_{0}^{2} \mathrm{V}_{\mathrm{eff}}},
\label{eq2}
\end{equation}
with
\begin{equation}
\mathrm{V}_{\mathrm{eff}}=\int_{V} \varepsilon(r)|\Phi(r)|^{2} d V,
\label{eq3}
\end{equation}
where $c$ represent the speed of light in vacuum, $n_{0}$ ($n_{2}$) are the linear (nonlinear) refractive index of the material with ranges $2 \leqslant n_{0} \leqslant 4$ and $10^{-13} \geq n_{2} \geq 10^{-17} \mathrm{cm}^{2} / \mathrm{W}$~\cite{bookjj}. $\mathrm{V}_{\text {eff }}$ is defined as the effective mode volume and describes the peak electric field strength within the cavity. $\varepsilon(r)$ and $\Phi(r)$ are the dielectric constant and the electric field strength, respectively.   $\mathrm{V}_{\text {eff }}$ is  located in between $10^{2} \sim 10^{4} \mu \mathrm{m}^{3}$ when the quality factor of the cavity $Q$  is limited in $\left(10^{6} \sim 10^{8}\right)$~\cite{Vahala2003}. For a near-infrared wavelength ($\lambda=1064 \mathrm{nm}$), the  nonlinear Kerr coefficient $u$ is estimated on the order between $0.0006 \sim 2721\mathrm{Hz}$ in a silica microsphere by calculating Eq.~(\ref{eq3}) with experimentally accessible parameters.

We consider the full description of the systemic dynamics including the fluctuation-dissipation processes of the optical and the mechanical modes.  After transforming the cavity mode to a randomly rotating frame according to  $\hat{a} \rightarrow \hat{a}_{\mathrm{p}} e^{i \varphi(t)}$, we derive a set of quantum  Langevin equations governing the systemic dynamics    
\begin{equation}
\begin{split}
\dot{\hat{a}}_\text{p}=&-(i\Delta_0+\kappa)\hat{a}_\text{p}+ig_0\hat{a}_\text{p}(\hat{b}^{\dagger}+\hat{b})-i\dot{\varphi}(t)\hat{a}_\text{p}\\&
+E_L-2iu\hat{a}^{\dagger}\hat{a}^2+\sqrt{2\kappa}\hat{a}_{\text{p,in}},\\
\dot{\hat{b}}=&-(i\omega_m+\gamma_m)\hat{b}+ig_0\hat{a}^{\dagger}_\text{p}\hat{a}_\text{p}
+2i\eta (\!\hat{b}+\hat{b}^{\dagger})^3
\\&+\sqrt{2\gamma_m}\hat{b}_{\text{in}},
\end{split}
\label{eq4}
\end{equation} 
where $\gamma_m$ describes the damping rate of the mechanical mode,  $\hat{b}_{\text{in}}$ is the thermal noise operator acting on the mechanical oscillator, and $\hat{a}_{\text{p,in}}$ is the squeezed input vacuum noise operator. The squeezed input optical field has been used to enhance sideband cooling and suppress Stokes scattering~\cite{PhysRevA.94.051801,Clark2017}. The corresponding noise correlations are  written  as
\begin{equation}
\begin{split}
\left\langle \hat{a}_{\mathrm{p,in}}(t)  \hat{a}_{\mathrm{p,in}}^{\dagger}\left(t^{\prime}\right)\right\rangle &=(N+1) \delta\left(t-t^{\prime}\right), \\
\left\langle  \hat{a}_{\mathrm{p,in}}^{\dagger}(t)  \hat{a}_{\mathrm{p,in}}\left(t^{\prime}\right)\right\rangle &=N \delta\left(t-t^{\prime}\right), \\
\left\langle  \hat{a}_{\mathrm{p,in}}(t)  \hat{a}_{\mathrm{p,in}}\left(t^{\prime}\right)\right\rangle &=M \delta\left(t-t^{\prime}\right), \\
\left\langle \hat{b}_{\text{in}}(t) \hat{b}^{\dagger}_{\text{in}}\left(t^{\prime}\right)\right\rangle &=(N_{\text{th}}+1) \delta\left(t-t^{\prime}\right),  \\
\left\langle \hat{b}^{\dagger}_{\text{in}}(t) \hat{b}_{\text{in}}\left(t^{\prime}\right)\right\rangle &=N_{\text{th}} \delta\left(t-t^{\prime}\right) ,
\end{split}
\label{eq5}
\end{equation}
where  $N=\sinh ^{2}\left(r_{e}\right)$ is the mean photon number of the broadband-squeezed vacuum environment, $M=\sinh \left(r_{e}\right) \cosh \left(r_{e}\right) e^{-i \Phi_{e}}$ is the strength of the autocorrelation of the squeezed vacuum  noise, and $N_{\text{th}}=1/\left(\exp \left\{\hbar \omega_{\mathrm{m}} / k_{\mathrm{B}} T\right\}-1\right)$ is the equilibrium mean thermal photon number. Here $r_{e}$ ($\Phi_{e}$) are the squeezing amplitude (angle) of the broadband-squeezed vacuum environment, $k_{\mathrm{B}}$ is the Boltzmann constant, and $T$ is the temperature of the mechanical oscillator. The system is linearized by substituting operators $\hat{a}_{\mathrm{p}}=\alpha+\delta \hat{a}_{\mathrm{p}}$ and $\hat{b}=\beta+\delta \hat{b}$ to Eq.~(\ref{eq4}) where $\alpha$ ($\beta$) and $\delta\hat{a}$ ($\delta\hat{b}$) describe the mean values and the fluctuations of the optical (mechanical) mode. Therefore, we can simplify the linearized Langevin equation as 
\begin{equation}
\begin{split}
\dot{\delta\hat{a}}_\text{p}=&-(i\Delta+\kappa)\delta\hat{a}_{\text{p}}+ig_0\alpha(\delta\hat{b}^{\dagger}+\delta\hat{b})-i\dot{\varphi}(t)\alpha\\
&-i\Omega\delta\hat{a}^{\dagger}_{\text{p}}+\sqrt{2\kappa}\hat{a}_{\text{p,in}}, \\
\dot{\delta\hat{b}}=&-(\gamma_m+i\omega^{\prime}_{m})\delta\hat{b}+ig_0(\alpha^*\delta\hat{a}_p+\alpha\delta\hat{a}^{\dagger}_p) 
\\&+i\Omega_m\hat{b}^{\dagger}+\sqrt{2\gamma}\hat{b}_{\text{in}},
\end{split}
\label{eq6}
\end{equation}
where we have ignored the higher-order nonlinear terms, and the effective parameters are listed as: $\Delta=\Delta_0-2g_0\Re{(\beta)}+4u\vert\alpha\vert^2$; $\Omega=2u\alpha^2$; $\omega^{\prime}_m=\omega_m-\Omega_m$; $\Omega_m=6\eta(4\Re{(\beta)}^2+1)$.  The mean values of the optical and the mechanical modes can be obtained by solving the steady Langevin equation
 \begin{equation}
\begin{split}
&-(i(\Delta-2u\vert\alpha\vert^2)+\kappa)\alpha+E_L=0, \\
&\!-(i\omega_{m}\!+\!\gamma_m)\beta\!+\!ig_0\vert\alpha\vert^2\!+\!i4\eta(4\Re{(\beta)}^3\!+\!3\Re{(\beta)})\!=0.
\end{split}
\label{eq7}
\end{equation}
We rewritten the strength of  parametric  amplification coefficient as $\Omega=\vert\Omega\vert e^{-2i\theta}$ (i.e., $\alpha=\vert\alpha\vert e^{-i\theta}$) with real angle $\theta$. The phase factors can be absorbed into the operators $\hat{a}_{\text{p}}$ (i.e., $\hat{a}_{\text{p}}\rightarrow\hat{a}_{\text{p}}e^{-i\theta}$). Therefore, we can obtain the  linearized Hamiltonian 
\begin{equation}
\begin{split}
\hat{H}_{\text{lin}}=&\Delta\delta\hat{a}^{\dagger}_{\text{p}}\delta\hat{a}_{\text{p}}+\omega^{\prime}_{m}\delta\hat{b}^{\dagger}\delta\hat{b}-g(\delta\hat{a}_{\text{p}}\!+\!\delta\hat{a}^{\dagger}_{\text{p}})(\delta\hat{b}+\delta\hat{b}^{\dagger})
\\
&+\frac{\vert\Omega\vert}{2}(\delta\hat{a}^{\dagger 2}_{\text{p}}+\delta\hat{a}^2_{\text{p}})
-\frac{\Omega_m}{2}(\delta\hat{b}^{\dagger 2}+\delta\hat{b}^2),
\end{split}
\label{eq8}
\end{equation}
where we have defined the optomechanical coupling $g$ as $g=g_0\vert\alpha\vert$. It is obvious that the Kerr nonlinear medium and the Duffing nonlinearity lead to the optical and the mechanical parametric amplification terms with amplitudes $\vert\Omega\vert$ and $\Omega_m$, respectively. We exploit squeezing transformations  $\delta\hat{a}_\text{p}=\cosh{(r)}\delta\hat{a}_\text{s}-\sinh{(r)}\delta\hat{a}^{\dagger}_\text{s}$ and $\delta\hat{b}=\cosh{(r_m)}\delta\hat{b}_\text{s}+\sinh{(r_m)}\delta\hat{b}^{\dagger}_\text{s}$ acting on the linearized Hamiltonian $\hat{H}_{\text{lin}}$ where the squeezing phase is fixed as $\pi$.  Here we choose the squeezing strength $r=\frac{1}{4}\ln(\frac{1+\eta}{1-\eta})$ and $r_m=\frac{1}{4}\ln(\frac{1+\eta_1}{1-\eta_1})$ with $\eta=\vert\Omega\vert/\Delta$  and $\eta_1=\vert\Omega_m\vert/\omega^{\prime}_m$.   Therefore, we can obtain the following effective Hamiltonian 
\begin{equation}
\hat{H}_{\text{e}}=\Delta_{\text{e}}\delta\hat{a}^{\dagger}_{\text{s}}\delta\hat{a}_{\text{s}}+\Delta_{m}\delta\hat{b}^{\dagger}_s\delta\hat{b}_s-G(\delta\hat{a}_{\text{s}}+\delta\hat{a}^{\dagger}_{\text{s}})(\delta\hat{b}^{\dagger}_s+\delta\hat{b}_s),
\label{eq9}
\end{equation}
where the effective coupling strength is $G=g\exp{(r^{\prime}})$ with $r^{\prime}=r_m-r$, and the effective detuning of the optical and mechanical modes are $\Delta_{\text{e}}=\Delta\sqrt{1-\eta^2}$ and $\Delta_{\text{m}}=\omega^{\prime}_m\sqrt{1-\eta^2_1}$.  We adjust the squeezing amplitudes to satisfy  $r_m\geq r$, and thus the effective $G$ can remain a large value.

Then we discuss the statistical properties of laser phase noise. As shown in Eq.~(\ref{eq6}), we note that the laser phase noise affects the systemic dynamics by the additional noise term $-i\dot{\varphi}(t)\alpha$ and the influence of phase noise is mainly depended on the mean value of the cavity field $\alpha$. Generally, the single-photon coupling $g_0$ is very small and one need to improve the effective optomechanical coupling via a large $\alpha$. It is a terrible contradiction that the large mean value of cavity field  $\alpha$ will lead to a sizeable effective optomechanical coupling and phase noise when $\alpha$ is  huge. Therefore, it is significant to enhance the effective optomechanical coupling and suppress the influence of phase noise at the same time.  If the phase noise correlation satisfies  $\langle\dot{\varphi}(t) \dot{\varphi}(t)\rangle=2 \Gamma_{L} \delta(t-t^{\prime})$,  the  spectrum of the noise is flat and the cut-off frequency $\gamma_c\rightarrow\infty$ (i.e., $\mathcal{S}_{\dot{\varphi}}(\omega)=2 \Gamma_{L}$), where $\Gamma_{L}$ is the linewidth of the driving laser. However, the spectral density of the phase noise is not a flat spectrum due to the finite non-zero correlation time of phase noise. In other words, it is a finite bandwidth color noise. Generally,  the noise spectrum is equivalent to a low pass filtered white noise with the following spectrum and correlation function
~\cite{PhysRevA.80.063819,PhysRevA.91.033828,PhysRevA.98.053841}
\begin{equation}
S_{\dot{\varphi}}(\omega)=\frac{2\Gamma_{L}}{1+\frac{\omega^{2}}{\gamma_{c}^{2}}}, ~ ~ \langle\dot{\varphi}(t) \dot{\varphi}(t^{\prime})\rangle=\Gamma_{L} \gamma_{c} e^{-\gamma_{c}|t-t^{\prime}|},
\label{eq10}
\end{equation}
where $1/\gamma_{c}$ is correlation time of the laser phase noise so that the phase noise is suppressed at frequencies $\omega>\gamma_{c} $.  The correlation time decreases and the frequency noise starts reaching the white noise with the increasing of $\gamma_{c}$. Moreover, the frequency spectrum in (\ref{eq10}) is equivalent to the differential equation $\ddot{\varphi}(t)+\gamma_{c} \dot{\varphi}(t)=\varepsilon(t)$ where $\varepsilon(t)$ is a Gaussian random variable with the noise correlation function 
\begin{equation}
\langle\varepsilon(t) \varepsilon(t^{\prime})\rangle=2 \gamma_{c}^{2} \Gamma_{L} \delta(t-t^{\prime}).
\label{eq11}
\end{equation}
 We redefine an additional noise operator $\psi=\dot{\varphi}$ where $\psi$ satisfies the following differential equation
\begin{equation}
\dot{\psi}(t)+\gamma_{c} \psi(t)=\varepsilon(t).
\label{eq12}
\end{equation} 
Therefore, we can rewrite the Langevin equation with the quadrature fluctuations of  the optical field and the mechanical oscillator: $\hat{X}=(\delta\hat{a}_\text{s}+\delta\hat{a}^{\dagger}_{\text{s}})/\sqrt{2}$; $\hat{P}=(\delta\hat{a}_\text{s}-\delta\hat{a}^{\dagger}_{\text{s}})/{\sqrt{2}i}$; $\hat{X}_{m}=(\delta\hat{b}_s+\delta\hat{b}^{\dagger}_s)/\sqrt{2}$; $\hat{P}_m=(\delta\hat{b}_s-\delta\hat{b}^{\dagger}_s)/\sqrt{2}i$. Moreover, the  corresponding  input noise operators are amended as $\hat{X}^{\text{in}}=e^{r}(\hat{\tilde{a}}_{\text{p,in}}+\hat{\tilde{a}}^{\dagger}_{\text{p,in}})/\sqrt{2}$, $\hat{P}^{\text{in}}=e^{-r}(\hat{\tilde{a}}_{\text{p,in}}-\hat{\tilde{a}}^{\dagger}_{\text{p,in}})/\sqrt{2}i$, $\hat{X}^{\text{in}}_{m}=e^{-r_m}(\hat{b}_{\text{in}}+\hat{b}^{\dagger}_{\text{in}})/\sqrt{2}$ and $\hat{P}^{\text{in}}_m=e^{r_m}(\hat{b}_{\text{in}}-\hat{b}^{\dagger}_{\text{in}})/\sqrt{2}i$ with $\hat{\tilde{a}}_{\text{p,in}}=\hat{a}_{\text{p,in}}e^{i\theta}$.  We note that noise $\hat{X}^{\text{in}}$ and $\hat{P}^{\text{in}}$ own exponential factors, which means decreasing laser phase noise is accomplished by increasing thermal noise.  To suppress the increased thermal noise around the cavity field, we adjust the amplitude and phase of the squeezed vacuum environment. If the squeezing parameters satisfy the  conditions $r=r_e$ and $\Phi_e-2\theta=\pi$, the effective input noise of the cavity is equivalent to a vacuum noise and we can obtain the following noise correlation function
\begin{equation}
\begin{split}
\left\langle\hat{X}^{\mathrm{in}}(t) \hat{X}^{\mathrm{in}}\left(t^{\prime}\right)\right\rangle=\frac{1}{2}\delta(t-t^{\prime}), \\
\langle\hat{X}^{\text{in}}(t)\hat{Y}^{\text{in}}(t^{\prime})\rangle=-\frac{1}{2i}\delta(t-t^{\prime}), \\
\left\langle\hat{X}^{\mathrm{in}}(t) \hat{X}^{\mathrm{in}}\left(t^{\prime}\right)\right\rangle=\frac{1}{2}\delta(t-t^{\prime}), \\
\langle\hat{Y}^{\text{in}}(t)\hat{X}^{\text{in}}(t^{\prime})\rangle=\frac{1}{2i}\delta(t-t^{\prime}).
\end{split}
\label{eqrr}
\end{equation}
We derive these noise correlations in detail in the appendix \ref{Appendix1}. By combining Eqs. (\ref{eq9}), (\ref{eq12}) and (\ref{eqrr}), we can derive the Langevin equation to describe the dynamic evolution of the system. We rewrite the Langevin equation as a compact matrix form
\begin{equation}
\dot{\vec{u}}(t)=A\vec{u}(t)+\vec{n}(t),
\label{eq13}
\end{equation}
 where we have defined the vector of  continuous variable  fluctuation operators
$\vec{u}(t)=[\delta\hat{X}, \delta\hat{P}, \delta\hat{X}_m, \delta\hat{P}_m, \psi]^{\top}$, 
 and the corresponding  input noise vector   is 
$\vec{n}=[\sqrt{2\kappa}\hat{X}^{\text{in}}, \sqrt{2\kappa}\hat{P}^{\text{in}}, \sqrt{2\gamma_m}\hat{X}^{\text{in}}_{m}, \sqrt{2\gamma_m}\hat{P}^{\text{in}}_{m}, \varepsilon]^{T}$.  Moreover, the drift matrix is the 5$\times$5 matrix
\begin{equation}
\begin{split}
A = \left(\begin{array}{ccccc}
-\kappa & \Delta_{\text{e}} & 0 & 0 & 0\\
-\Delta_{\text{e}} &  -\kappa & 2ge^{r^{\prime}} & 0 & -\sqrt{2}\vert\alpha\vert e^{-r}\\
0 & 0 & -\gamma_m & \Delta_{m} & 0\\
2ge^{r^{\prime}} & 0 & -\Delta_{m} & -\gamma_m & 0 \\
 0 &  0 &  0 &  0 &  -\gamma_{c}
\end{array} \right),
\end{split}
\label{eq14}
\end{equation}
where the element $-\sqrt{2}\vert\alpha\vert e^{-r}$ in the drift matrix $A$ describes the coupling between the phase noise operator and the optical  momentum operator. It is different from the standard cavity optomechanical system that the effective coupling and the phase noise term multiply exponential factors $e^{r^{\prime}}$ and $e^{-r}$, respectively. One can enlarge the squeezing strength $r$ to suppress the influence of phase noise and  increase $r^{\prime}$ to improve the coupling $G$. Therefore,  the parametric processes, induced by the Kerr medium and Duffing nonlinearity, can simultaneously increase the effective optomechanical coupling and reduce the coupling between the laser phase noise and the cavity field. According to the Eq.~(\ref{eq13}), we obtain the following dynamical equation of the covariance matrix 
\begin{equation}
\frac{dV}{dt}=AV+VA^{T}+N,
\label{eq15}
\end{equation}
where the matrix element of  the  covariance matrix $V$ can be expressed as
\begin{equation}
  V_{ij}=\frac{1}{2}\left\langle \vec{u}_{i} \vec{u}_{j}+\vec{u}_{j} \vec{u}_{i}\right\rangle,
  \end{equation} 
 and the corresponding noise matrix is  
 \begin{equation}
\!N\!=\!\begin{pmatrix}
           \kappa &  0 & 0 & 0 & 0\\
0 & \kappa & 0 & 0 & 0\\
0 &  0 & \gamma_m\lambda(2n_{\text{th}}+1) & 0 & 0\\
0 & 0 & 0 & \frac{\gamma_m}{\lambda}(2n_{\text{th}}+1) & 0 \\
 0 &  0  &  0 &  0 &  2\gamma_{c}^{2} \Gamma_{L}
\end{pmatrix},
\label{eq16}
\end{equation}
 where we have defined the parameter $\lambda=e^{-2r_m}$. Generally, one exploits a squeezed vacuum bath to counteract the influence of the factor $\lambda=e^{-2r_m}$. However, the mechanical decay $\gamma_m$ is very small so that the factor $\lambda=e^{-2r_m}$ in Eq.~(\ref{eq16}) has a little affect on the systemic dynamics. Therefore, we retain this factor in the following calculations.
 
 \begin{figure}[tbp]
\centering
{\includegraphics[width=0.8\columnwidth]{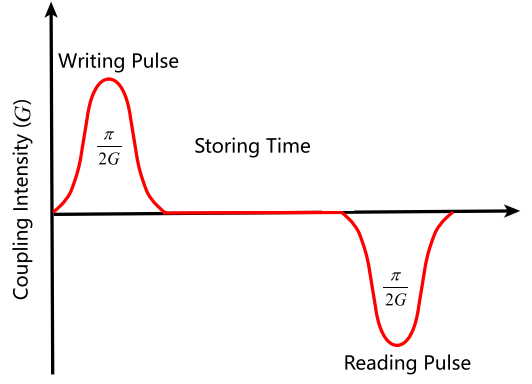}}
\caption{
 Firstly, a writing pulse steers the initial state of the cavity mode to the mechanical oscillator, which is the so-called writing process.  Then the state is stored in the mechanical oscillator for a while $\tau$.  The state is stored in the mechanical oscillator with a high fidelity due to a small damping rate.  Finally, a reading pulse is injected into the system with an opposite sign of the writing pulse for reading the stored state.}
\label{figure2}
\end{figure}
\section{Demonstrating some actual quantum phenomenon}
\label{Demonstrating some actual quantum phenomenon}
In this section, we take two examples to test the efficiency of our scheme mentioned in the above section. Firstly, we theoretically investigate the performance of our proposal on improving the optomechanical quantum memories against the laser phase noise.  Secondly, we exploit our design to demonstrate the stationary optomechanical entanglement.
\subsection{Quantum  memory}
\label{Quantum  memory}
\begin{figure}[tbp]
\centering
{\includegraphics[width=0.9\columnwidth]{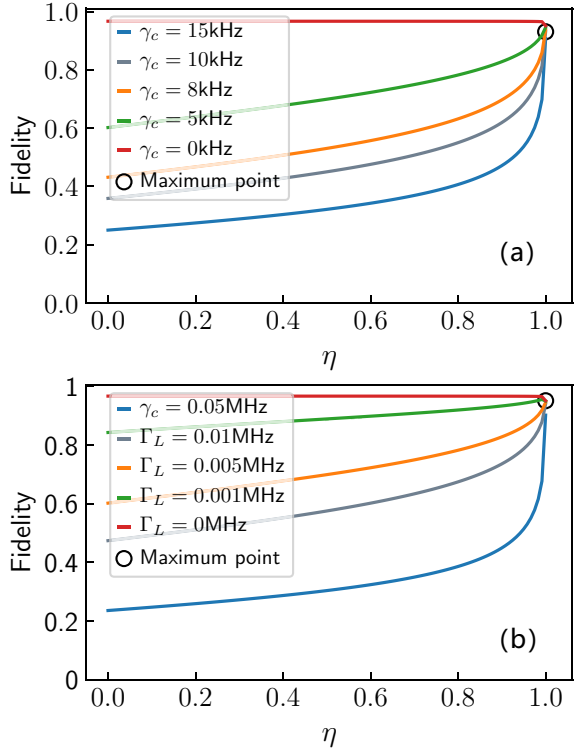}}
\caption{(Color~online) 
The fidelity of quantum memory is calculated as a function of parameter $\eta$ for different parameters related to phase noise: (a) the linewidth of driving laser $\Gamma_L=5$kHz;  (b) cut-off frequencies  $\gamma_c=5$kHz. Other parameters are given in the text.}
\label{figure3}
\end{figure}
\begin{figure}[tbp]
\centering
{\includegraphics[width=0.9\columnwidth]{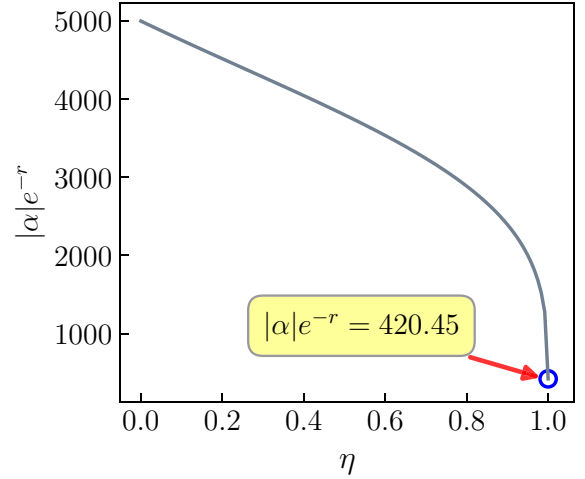}}
\caption{(Color~online) We simulate the mean value of the cavity field $\alpha$ as a function of $\eta$ by fixing $G=0.05\omega_m=3.14$MHz and $g=2\pi\times 100$Hz. $\eta$ is limited in $0\sim0.9999$. Other parameters are given in the text.}
\label{figure4}
\end{figure}
\begin{figure*}[tbp]
\centering
{\includegraphics[width=2\columnwidth]{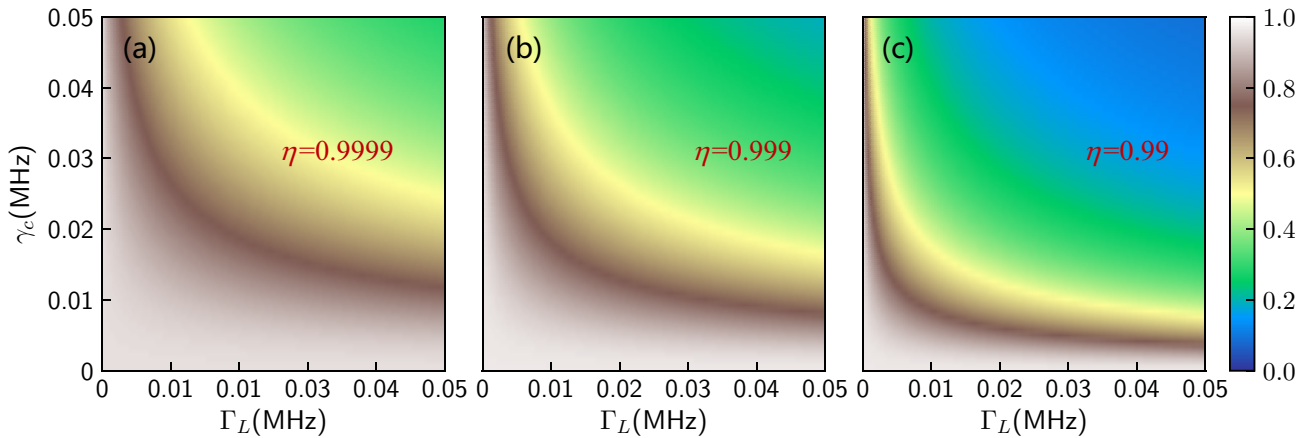}}
\caption{(Color~online) 
The fidelity of quantum memory is simulated as a function of  cut-off frequency $\gamma_c$ and laser linewidth  $\Gamma_L$ for different $\eta$:  (a)  $\eta=0.9999$;  (b) $\eta=0.999$; (c) $\eta=0.99$. Other parameters are given in the text.}
\label{figure5}
\end{figure*}

\begin{figure}[tbp]
\centering
{\includegraphics[width=0.9\columnwidth]{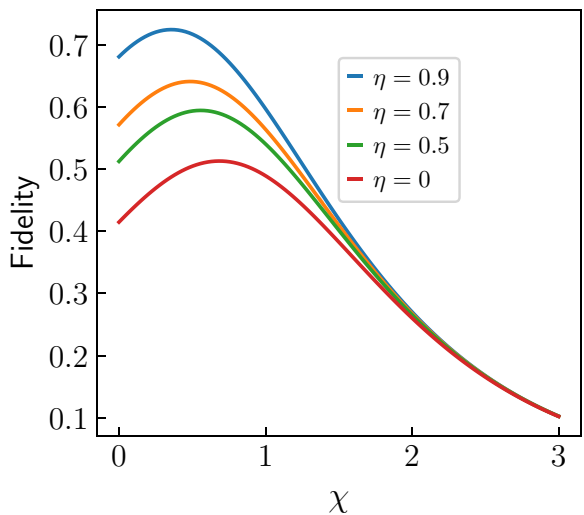}}
\caption{(Color~online) We simulate the fidelity as a function of the squeezing parameter $\chi$ for different $\eta$. Other parameters are given in the text.
}
\label{figure6}
\end{figure}
Quantum memory is indispensable for quantum information processing and has made enormous progress in optics and atoms~\cite{Felinto2006}. Let us  briefly recall the quantum memory in optomechanical system~\cite{PhysRevLett.107.133601}. As shown in Fig.~\ref{figure2},  the state of the optical mode is transferred to the mechanical mode in the writing process with time $\frac{\pi}{2G}$. Then the state is stored in the mechanical membrane for a while $\tau$ by decoupling the mechanical and the optical modes. Finally, the optical mode obtains the storied state in the reading process, and the corresponding reading time is $\frac{\pi}{2G}$. One of the advantages of quantum memory in optomechanical systems is that the decay rate of the mechanical oscillator is much smaller than the optical cavity.

Here we apply the model to achieve quantum memory. We suppose the effective detuning satisfies the resonate condition $\Delta_e\approx\Delta_m\gg G$ which can be achieved by controlling the frequency of the optical driving. This condition limits that $\eta$ cannot close to one infinitely. In other words, we cannot fully cancel the laser phase noise. In our paper, we limit $\eta$ in $0\sim 0.9999$. $F=\operatorname{Tr}\left(\hat{\rho}_{\mathrm{i}} \hat{\rho}_{\mathrm{f}}\right)$ is defined as the fidelity between the initial state $\hat{\rho}_{\mathrm{i}}$ and final state $\hat{\rho}_{\text{f}}$. In the phase space, one can rewrite the fidelity as $F=\pi \int_{-\infty}^{\infty} d \vec{\xi} W_{\text{i}}(\xi) W_{\text{f}}(\xi)$, where $\xi \in \Re^{2}$ is the vector of
the optical quadratures $\xi=\left[\delta\hat{X}, \delta\hat{P}\right]^{T}$ and  $W_{\text{i}}$ ($W_{\text{f}}$) are the Wigner functions of the optical initial (final) states.  To simplify the calculation, we assume the state of the cavity staying in a pure Gaussian state—the Wigner function $W_{\text{i}}$ ($W_{\text{f}}$) are the Gaussian distribution. Under this assumption, one  can obtain the fidelity between the initial and the final states of the optical mode at time
\begin{equation}
F=\frac{1}{1+\bar{n}_{\mathrm{h}}} \exp \left(-\frac{\Theta^{2}}{1+\bar{n}_{\mathrm{h}}}\right),
\label{eq17}
\end{equation}
where the parameters $\bar{n}_{\mathrm{h}}$, $\Theta$ has the following forms
\begin{equation}
\begin{array}{l}
\bar{n}_{\mathrm{h}}=2 \sqrt{\operatorname{det}(\frac{V_{\mathrm{i}}+V_{\mathrm{f}}}{2})}-1, \\
\Theta^{2}=(\langle\hat{\xi}_{\mathrm{i}}\rangle-\langle\hat{\xi}_{\mathrm{f}}\rangle) \cdot \frac{\sqrt{\operatorname{det}(\frac{V_{\mathrm{i}}+V_{\mathrm{f}}}{2})}}{V_{\mathrm{i}}+V_{\mathrm{f}}}(\langle\hat{\xi}_{\mathrm{i}}\rangle-\langle\hat{\xi}_{\mathrm{f}}\rangle),
\end{array}
\label{eq18}
\end{equation}
with the initial (final) covariance matrix $V_{\text{i}}$ ($V_{\text{f}}$) and the initial (final) optical mean quadratures $\xi_{\text{i}}$ ($\xi_{\text{f}}$). 	The detailed derivation of Eq.~(\ref{eq17}) have been proposed by Wang~\cite{Wang_2012}.
 
To measure the fidelity $F$, we need to calculate the expectations of optical quadratures ($\xi_{\text{i}}$ and $\xi_{\text{f}}$) and the covariance matrix ($V_{\text{i}}$ and $V_{\text{f}}$) of the initial and final states.  Here we consider the squeezed coherent state (a typical Gaussian state) as  the initial state of the optical field in the squeezing frame (i.e. $|\mu, \chi\rangle=D(\mu) S(\chi)|0\rangle$), and thus the state in the original frame is $S(r)^{\dagger}D(\mu) S(\chi)|0\rangle$ where $D(\mu)=\exp{(\mu \hat{a}^{\dagger}-\mu^{*} \hat{a})}$ and $S(\chi)=\exp{(\frac{\chi^*}{2}\hat{a}^{2}-\frac{\chi}{2}\hat{a}^{\dagger 2})}$ are displacement and squeezing operators, respectively.   Therefore, one can obtain the initial vector $\vec{u}(0)=[\sqrt{2}\operatorname{Re}(\mu), \sqrt{2}\operatorname{Im}(\mu), 0, 0, 0]^{T}$ and the corresponding initial covariance matrix is  
\begin{equation}
\begin{split}
V(0)=\frac{1}{2}\left(\begin{array}{ccccc}
e^{-2\chi} & 0 & 0 & 0 & 0\\
0 & e^{2\chi} & 0 & 0 & 0\\
0 &  0 & 1 & 0 & 0\\
0 & 0 & 0 & 1 & 0 \\
 0 &  0 &  0 &  0 &  1
\end{array} \right),
\end{split}
\label{eq181}
\end{equation}
where we have assumed $\chi\in\Re$.

To numerically analyze the quantum memory effect, we choose the parameters similar to those in Ref.~\cite{PhysRevA.84.032325,PhysRevA.84.063827}:  length of the cavity $L=1$mm; wavelength of the cavity field $1064$ nm; mass of the mechanical oscillator $m\simeq$10ng; frequency of the mechanical oscillator $\omega_{m}=2\pi \times 10\mathrm{MHz}$; quality factor $Q_{m}=2\times 10^6$; the optical decay rate  $\kappa=2\pi\times 100$kHz; single-photon coupling strength  $g_0=2\pi\times 100$Hz; the linewidth of driving laser $\Gamma_L$ in range $1\sim 50$kHz;  the cut-off frequency $\gamma_c$ in range $0.1\sim 50$kHz. Moreover, we assume the mean thermal phonon number $n_{\text{th}}=3$ and the parameter $r^{\prime}=0$. The storage time is $\tau=65\omega^{-1}_m=1.035\mu s$.  We fix the effective coupling $G=0.05\omega_m$ which can be achieved by controlling the strength of driving laser. 

To demonstrate the advantage of  our proposal, we simulate the fidelity $F$ as a function of parameter $\eta$ in Figs.~\ref{figure3}(a) and \ref{figure3}(b) for different cut-off frequency $\gamma_c$ and laser linewidth $\Gamma_L$, respectively, where we have limited the parameter $\eta$ in interval $[0, 0.9999]$.   It is clear that the fidelity is improving with the increasing of parameter $\eta$ even the noise spectrum has large cut-off frequency $\gamma_c$ and laser linewidth  $\Gamma_L$.  In particular, the fidelity is approximately the same with ideal situation (i.e., without laser phase noise $\gamma_c=0$)  for $\eta=0.9999$ though the cut-off frequency $\gamma_c$ takes the values $5$kHz, $8$kHz, $10$kHz, and $15$kHz. It indicates that the phase noise is extremely suppressed even can be approximately ignored, which is  largely different from the case $\eta=0$ (the standard optomechanical system). We improve the parameter $\eta$ to increase the squeezing parameter $r$ (i.e., enhancing $e^r$) so that  the noise term $-\sqrt{2}\vert\alpha\vert e^{-r}\dot{\varphi}(t)$ can be effectively suppressed. Moreover, we numerically simulate  the variation of $\vert\alpha\vert$ with $\eta$ in Fig.~\ref{figure4}. It is obvious that $\vert\alpha\vert$ is monotonically decreasing with the increasing of parameter $\eta$ and the minimum value $\vert\alpha\vert e^{-r}$ is $420.45$ for $\eta=0.9999$. At this time, the effect of phase noise is reduced about an order of magnitude.

 To further clarify the promotion effect of large $\eta$ (i.e., large $r$), we simulate the fidelity as the function of cut-off frequency  $\gamma_c$ and laser linewidth $\Gamma_L$ for different $\eta$ in Figs.~\ref{figure5}(a)-\ref{figure5}(c). The results show that the area of high fidelity shrinks with the decreasing of  $\eta$, and the destructive effect of  the cut-off frequency $\gamma_c$ on the fidelity $F$ is more larger than the laser linewidth $\Gamma_L$. Moreover, the fidelity can arrive at 0.955 for $\eta=0.9999$ even the parameters of  the noise spectrum are very huge ($\gamma_c=0.01$MHz and $\Gamma_L=0.01$MHz). In Fig.~\ref{figure6}, we simulate the fidelity $F$ versus to the increasing of the squeezing amplitude of the initial state which is described by the squeezing parameters $\chi$. One can easily find the fidelity decreases by increasing parameter $\chi$, while the variation of the fidelity for $\eta=0.9$ is slower than  $\eta=0.7$, $\eta=0.5$, and $\eta=0$. Therefore, our scheme can protect the fidelity and inhibit the phase noise for a small $\chi$. However, the advantage of the scheme slowly disappears and the initial state would be more sensitive to various noises when the initial state becomes more and more non-classical (i.e., with the increasing of $\chi$).

\subsection{Entanglement}
\label{entanglement}
\begin{figure}[tbp]
\centering
{\includegraphics[width=0.9\columnwidth]{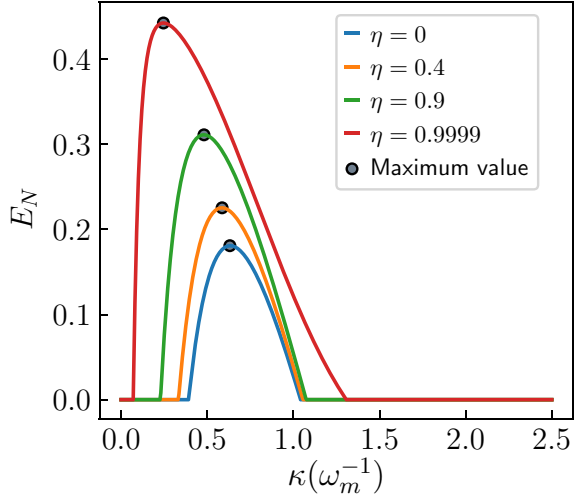}}
\caption{(Color~online) 
The stationary optomechanical entanglement $E_N$ versus the decay of  cavity mode for $\eta=0$, $\eta=0.4$, $\eta=0.9$, and $\eta=0.9999$.  The effective coupling is $G/\omega_m=0.5$, the cut-off frequency is $\gamma_c=10$kHz, and the laser linewidth is $\Gamma_L=10$kHz. Other parameters are given in the text.
}
\label{figure7}
\end{figure} 
 \begin{figure}[tbph]
\centering
{\includegraphics[width=0.9\columnwidth]{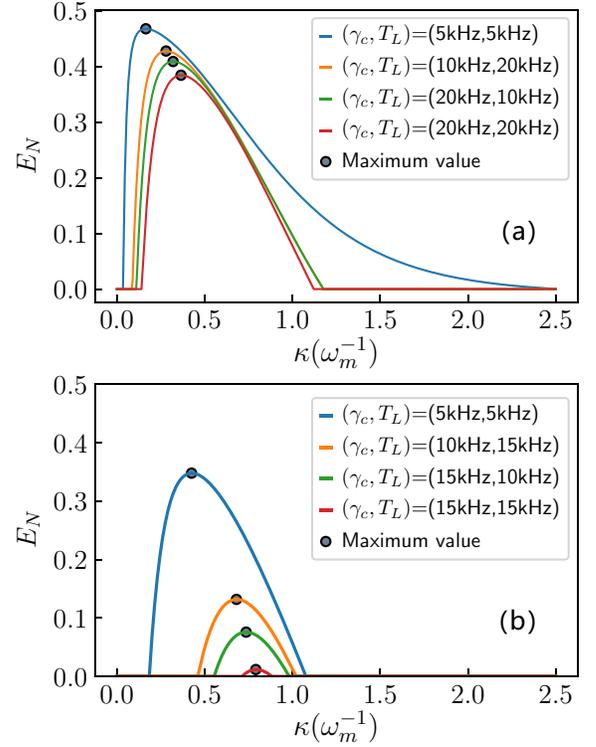}}
\caption{(Color~online) 
The stationary optomechanical entanglement $E_N$ versus to the decay $\kappa$ of the  cavity mode. (a) the parameter $\eta$ is 0.9999 under various cut-off frequencies $\gamma_c$ and  laser linewidth $\Gamma_L$ of the driving laser. (b) the parameter $\eta$ equals to 0 under various cut-off frequencies $\gamma_c$ and  laser linewidth $\Gamma_L$ of the driving laser, which indicates the standard optomechanical model ($r=0$).  Other parameters are the same with Fig.\ref{figure7}.
}
\label{figure8}
\end{figure}
\begin{figure*}[tbph]
\centering
{\includegraphics[width=2\columnwidth]{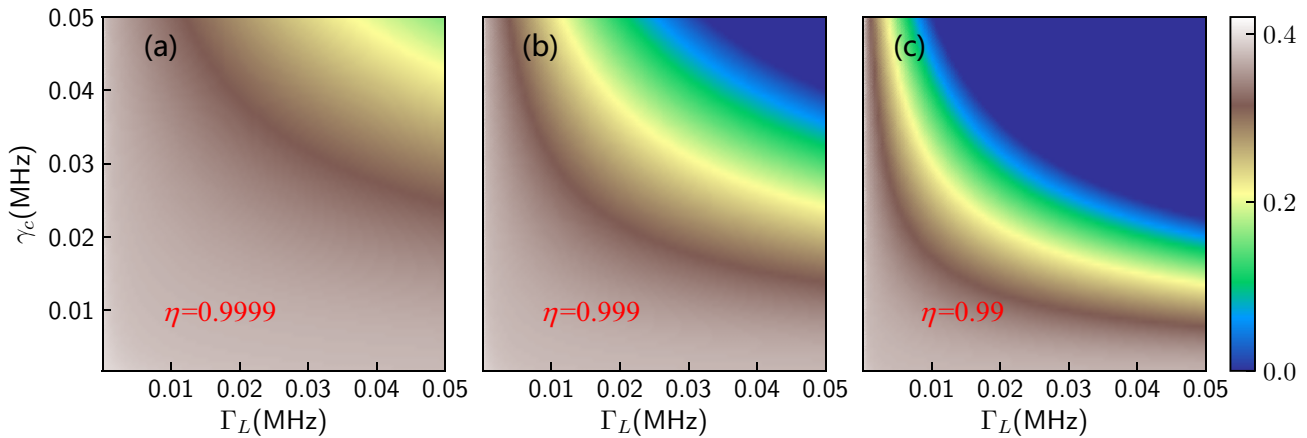}}
\caption{(Color~online)  Contour plot of the stationary optomechanical entanglement $E_N$
versus cut-off frequencies $\gamma_c$ and spectral widths of phase noise for (a) $\eta=0.9999$,  (b) $\eta=0.999$, and (c) $\eta=99$. The decay of the cavity is $\kappa/2\pi=5$MHz.  Other parameters are the same with Fig.\ref{figure7}.
}
\label{figure9}
\end{figure*}

 Here we consider the stationary optomechanical entanglement as the second example to demonstrate the advantage of our scheme on suppressing the phase noise. The steady-state of the system is associated with Eq.~(\ref{eq15}). If and only if all the eigenvalues of the matrix $A$ have negative real parts, the system can arrive in the steady-state. We derive the stability conditions by using the Routh-Hurwitz criteria~\cite{PhysRevA.35.5288,Mahajan2019652,Bhatt2019,Mahajan2013}. According to the criteria, we get the following  two non-trivial stability conditions: 
\begin{subequations}
\begin{equation}
\gamma _m^2 \Delta _e^2+\kappa ^2 \gamma _m^2+\kappa ^2 \Delta _m^2+(\Delta _e\Delta _m-4 G^2) \Delta _e \Delta _m >0,
\label{eq21a}
\end{equation}
\begin{equation}
\begin{split}
&4 \kappa\gamma _m\left\lbrace \left((\Delta _e-\Delta _m)^2+(\kappa+\gamma _m)^2\right)\left((\Delta _e+\Delta _m)^2\right. \right.\\&
\left.\left.+(\kappa+\gamma _m)^2\right)\right\rbrace+
16 G^2 \Delta _e \left(\kappa +\gamma _m\right){}^2 \Delta _m>0.
\end{split}
\label{eq21b}
\end{equation}
\end{subequations}
In the next calculation, we restrict all the parameters to satisfy the stable condition~(\ref{eq21a}) and (\ref{eq21b}). Here we consider the frequency condition $\Delta_e, \Delta_m>0$. It should be noticed that the critical condition (\ref{eq21a}) limits the exponential improvement of coupling and noise suppression (i.e, $0<r<r_{\text{max}}$) where
\begin{equation}
r_{\text{max}}=\frac{1}{2}\ln(\frac{\gamma_m^2 \Delta^2_e+\kappa^2 \gamma_m^2+\kappa^2 \Delta^2_m+\Delta^2 _e\Delta^2_m}{4g^2\Delta_e\Delta_m}).
\label{eq22new}
\end{equation}
 For our model, the steady-state is a zero-mean Gaussian state because we have linearized the dynamics of the fluctuations, and all noises are Gaussian; as a consequence, it is fully characterized by the $5 \times 5$ stationary covariance matrix  $V(\infty)$ with matrix elements,
\begin{equation}
V_{\text{ij}}=\frac{\left\langle u_{\text{i}}(\infty) u_{\text{j}}(\infty)+u_{\text{j}}(\infty) u_{\text{i}}(\infty)\right\rangle}{2}.
\label{eq20}
\end{equation}
The stationary covariance matrix $V(\infty)$ can be obtained  by solving the following Lyapunov equation
\begin{equation}
AV+V A^{\top}=-N.
\label{eq21}
\end{equation}
We find that the Lyapunov equation (\ref{eq21}) is linear for the covariance matrix $V$, which means the Lyapunov equation can be straightforwardly and analytically solved. One can identify all the quantum properties of the stationary state of the optomechanical system according to the stationary covariance matrix $V(\infty)$. Therefore, we can simulate the influence of laser phase noise on achieving quantum entanglement between the mechanical oscillator and optical mode. The stationary optomechanical entanglement relates to the mechanical and optical quadratures, and thus we concentrate on the reduced $4\times 4$ covariance matrix of $V(\infty)$. This reduced correlation matrix has the following form
\begin{equation}
V \equiv\left(\begin{array}{ll}
V_{\mathrm{A}} & V_{\mathrm{C}} \\
V_{\mathrm{C}}^{\mathrm{T}} & V_{\mathrm{B}}
\end{array}\right),
\label{eq22}
\end{equation}
where $V_{\mathrm{A}}, V_{\mathrm{B}}$ and $V_{\mathrm{C}}$ are $2 \times 2$ matrix. The  matrix $V_{\mathrm{A}}$ ($V_{\mathrm{B}}$)  are associated with the optical mode (mechanical oscillator), while $V_{\mathrm{C}}$ describes the optomechanical correlations. The logarithmic negativity is the famous and convenient measure for predicting  continuous variable (CV) entanglement \cite{PhysRevA.65.032314},  and its definition is
\begin{equation}
E_{N}=\max \left(0,-\ln 2 \eta^{-}\right)
\label{eq23}
\end{equation}
where $\eta^{-}$ is the symplectic eigenvalue of the bipartite system, and it has the following form 
\begin{equation}
\eta^{-} \equiv \frac{1}{\sqrt{2}}\left[\Sigma(V)-\sqrt{\Sigma(V)^{2}-4 \operatorname{det} V}\right]^{1 / 2}
\label{eq24}
\end{equation}
with $\Sigma(V)=\operatorname{det} V_{\mathrm{A}}+\operatorname{det} V_{\mathrm{B}}-2 \operatorname{det} V_{\mathrm{C}}$.

Then we study the advantage of  our scheme on CV entanglement when the phase noise exists in the system. In Figs.~\ref{figure7}(a) and (b), we  simulate $E_N$ versus to the optical decay $\kappa$ under different $\eta$. Obviously,  laser phase noise has a prominent effect on the stationary optomechanical entanglement while the large parameter $\eta$ can improve the maximum value of $E_N$ and broaden the parameter region existing entanglement. By comparing with Figs.~\ref{figure7}(a) and (b), we find  our proposal ($\eta=0.9999$) has a great advantage than the standard optomechanical coupling model ($\eta=0$).  Therefore, our scheme extremely inhibits the negative effect of  laser phase noise on the stationary optomechanical entanglement, which improve the conclusion in Ref. \cite{PhysRevA.84.032325}. Moreover, we simulate $E_N$ as a function of  $\kappa$ for $\eta=0.9999$ and $\eta=0$ in Figs.~\ref{figure8}(a) and \ref{figure8}(b), respectively. According to the numerical results, we summarize the advantages of our scheme versus the standard optomechanical coupling model: the entanglement $E_N$ is more greater for huge $\gamma_c$ and $\Gamma_L$; $E_N$ can exist in a wide range of $\kappa$; the entanglement $E_N$ decays more slowly with the increasing of  $\gamma_c$ and $\Gamma_L$; $E_N$ exists even for larger laser phase noise  $(\gamma_c, \Gamma_L)=(20\text{kHz}, 20\text{kHz})$. However, the standard optomechanical coupling model ($\eta=0$) is sensitive to laser phase noise, and the stationary optomechanical entanglement $E_N$ is approximately close to zero for $(\gamma_c, \Gamma_L)=(15\text{kHz}, 15\text{kHz})$. In Fig.~\ref{figure9}, we simulate $E_N$ versus the cut-off frequency $\gamma_c$ and the laser linewidth  $\Gamma_L$ for (a) $\eta=0.9999$, (b) $\eta=0.999$, and (c) $\eta=99$. The numerical results show that the destructive effect of phase noise is tiny when $\eta$ is closer to one. Although the maximum achievable entanglement decreases with the increasing of cut-off frequency $\gamma_c$ and  laser linewidth  $\Gamma_L$, we still obtain a large parameters range to maintain the entanglement for a large $\eta$. Moreover, we also notice that the destructive effect of  $\gamma_c$ is more remarkable than $\Gamma_L$. Therefore, our method can suppress the laser phase noise,  and thus protect the stationary entanglement.
\section{Conclusion}
\label{Conclusion}
In summary, we studied a theoretical proposal to suppress the phase noise and improve the effective optomechanical coupling.  The optomechanical system includes a mechanical Duffing nonlinearity and a Kerr medium that can create the mechanical and optical parametric amplification terms. Further calculation shows that we can enhance the effective optomechanical coupling and inhibit the laser phase noise at the same time. In this process, we use the squeezed vacuum environment to inhibit the increased thermal noise. To test the performance of our proposal, we simulate quantum memory and stationary optomechanical entanglement as examples. The numerical results show that our scheme effectively suppresses the destructive influence of laser phase noise on quantum memory and protects the storing fidelity at a high value. Moreover, our proposal can also protect stationary optomechanical entanglement. In particular, the maximal entanglement decreases very slowly with the increasing of the laser phase noise, and it exists in wide ranges of parameters. Our scheme provides a promising way for inhibiting the phase noise of optomechanical systems or other quantum systems driven by lasers and has potential applications for achieving quantum information processes and observing quantum phenomena.

\section*{ACKNOWLEDGMENTS}
The authors thank Wenlin Li, Feng-Yang Zhang, and Denghui Yu for the useful discussion. This research was supported by the National Natural Science Foundation of China (Grant No. 11574041 and 11375036) and the Excellent young and middle-aged Talents Project in scientific research of Hubei Provincial Department of Education (under Grant No. Q20202503)
\appendix
\renewcommand\thesection{\Alph{section}}
\setcounter{equation}{0}

\section{The effective noise correlation of the effective mode}
\label{Appendix1}
Here, we apply the method proposed in~\cite{PhysRevLett.114.093602,PhysRevA.95.053861,PhysRevA.91.013834} to suppress the increased thermal noise. By setting the phase and amplitude of the squeezed vacuum field, we suppress the correlations of the effective thermal noise that have the following forms
\begin{subequations}
\begin{equation}
\begin{split}
\langle\hat{X}^{\text{in}}(t)\hat{X}^{\text{in}}(t^{\prime})\rangle
=&\frac{e^{2r}}{2}(\sinh(r_e)^2+\cosh(r_e)^2
\\&+\sinh(2r_e)\cos(\Phi_e-2\theta))\delta(t-t^{\prime})\\
=&\frac{1}{2}\delta(t-t^{\prime}),
\end{split}
\label{aeq1}
\end{equation}
\begin{equation}
\begin{split}
 \langle\hat{Y}^{\text{in}}(t)\hat{Y}^{\text{in}}(t^{\prime})\rangle=&\frac{e^{2r}}{2}(\sinh(r_e)^2+\cosh(r_e)^2
\\
&-\sinh(2r_e)\cos(\Phi_e-2\theta))\delta(t-t^{\prime})\\
=&\frac{1}{2}\delta(t-t^{\prime}),
\end{split}
\end{equation}
 \begin{equation}
\begin{split}
\langle\hat{X}^{\text{in}}(t)\hat{Y}^{\text{in}}(t^{\prime})\rangle=&-\frac{1}{2i}(\sinh(r_e)^2-\cosh(r_e)^2
\\&-i\sinh(2r_e)\sin(\Phi_e-2\theta))\delta(t-t^{\prime})\\
=&-\frac{1}{2i}\delta(t-t^{\prime}),
\end{split}
\end{equation}
\begin{equation}
\begin{split}
\langle\hat{Y}^{\text{in}}(t)\hat{X}^{\text{in}}(t^{\prime})\rangle=&\frac{1}{2i}(\cosh(r_e)^2-\sinh(r_e)^2
\\&-i\sinh(2r_e)\sin(\Phi_e-2\theta))\delta(t-t^{\prime})\\
=&\frac{1}{2i}\delta(t-t^{\prime}),
\end{split}
\end{equation}
\end{subequations}
where we have supposed the condition $r=r_e$ and $\Phi_e-2\theta=\pi$. 



\bibliographystyle{apsrev4-1}
\bibliography{mybibtex}

\end{document}